\documentclass[twocolumn]{aastex631}

\usepackage{multirow}
\usepackage{amsmath}
\usepackage{soul, color, xcolor}
\usepackage{natbib}
\soulregister{\cite}{1}
\soulregister{\citealt}{1}
\usepackage{tabularx}
\usepackage{longtable}
\usepackage{array}
\usepackage{makecell}
\usepackage[T1]{fontenc}
\shorttitle{DAVOS: DES-SN}
\shortauthors{Liu et al.}

\begin{document}

\title{Dwarf Active Galactic Nuclei from Variability for the Origins of Seeds (DAVOS): Properties of Variability-Selected AGNs in the Dark Energy Survey Deep Fields}

\correspondingauthor{Colin J. Burke}
\email{colin.j.burke@yale.edu}

\author[0000-0003-4247-0169]{Yichen Liu}
\affiliation{Department of Astronomy, University of Illinois at Urbana-Champaign, 1002 W. Green Street, Urbana, IL 61801, USA}
\affiliation{National Center for Supercomputing Applications, University of Illinois at Urbana-Champaign, 605 East Springfield Avenue, Champaign, IL 61820, USA}
\affiliation{Steward Observatory, University of Arizona, 933 N Cherry Ave, Tucson, AZ 85719}

\author[0000-0001-9947-6911]{Colin J. Burke}
\affiliation{Department of Astronomy, Yale University, 219 Prospect Street, New Haven, CT 06511, USA}

\author[0009-0009-9486-3053]{Diego Miura}
\affiliation{Department of Astronomy, Yale University, 219 Prospect Street, New Haven, CT 06511, USA}

\author[0000-0003-0049-5210]{Xin Liu}
\affiliation{Department of Astronomy, University of Illinois at Urbana-Champaign, 1002 W. Green Street, Urbana, IL 61801, USA}
\affiliation{National Center for Supercomputing Applications, University of Illinois at Urbana-Champaign, 605 East Springfield Avenue, Champaign, IL 61820, USA}
\affiliation{Center for Artificial Intelligence Innovation, University of Illinois at Urbana-Champaign, 1205 West Clark Street, Urbana, IL 61801, USA}

\author[0000-0002-6893-3742]{Qian Yang}
\affiliation{Center for Astrophysics | Harvard \& Smithsonian, 60 Garden Street, Cambridge, MA 02138, USA}

\author[0000-0002-5554-8896]{Priyamvada Natarajan}
\affiliation{Department of Astronomy, Yale University, 219 Prospect Street, New Haven, CT 06511, USA}
\affiliation{Department of Physics, Yale University, 217 Prospect Street, New Haven, CT 06520, USA}
\affiliation{Black Hole Initiative, Harvard University, 20 Garden Street, Cambridge, MA 02138, USA}

\author[0000-0002-4557-6682]{Charlotte A. Ward}
\affiliation{Department of Astrophysical Sciences, Princeton University, Princeton, NJ 08544, USA}

\DeclareRobustCommand{\colin}[1]{{\sethlcolor{Yellow}\hl{(CJB:) #1}}}
\DeclareRobustCommand{\yichen}[1]{{\sethlcolor{lime}\hl{(YL:) #1}}}
\DeclareRobustCommand{\revision}[1]{{\sethlcolor{red}\hl{(rev:) #1}}}
\DeclareRobustCommand{\original}[1]{{{{\sethlcolor{lightgray}\hl{(original:)}}}\color{lightgray}{ #1}}}



\begin{abstract}


We study the black hole mass $-$ host galaxy stellar mass relation, $M_{\rm{BH}}-M_{\ast}$, for a sample of 706 $z \lesssim 1.5$ and $i \lesssim 24$ optically-variable active galactic nuclei (AGNs) in three Dark Energy Survey (DES) deep fields: C3, X3, E2, which partially cover Chandra Deep Field-South, XMM Large Scale Structure survey, and European Large Area ISO Survey, respectively. The parent sample was identified by optical variability from the DES supernova survey program imaging. Using publicly available spectra and photometric catalogs, we consolidate their spectroscopic redshifts, estimate their black hole masses using broad line widths and luminosities, and obtain improved stellar masses using spectral energy distribution fitting from X-ray to mid-infrared wavelengths. Our results confirm previous work from Hyper-Suprime Camera imaging that variability searches with deep, high-precision photometry can reliably identify AGNs in low-mass galaxies up to $z\sim1$. However, we find that the hosted black holes are overmassive than predicted by the local AGN relation, fixing host galaxy stellar mass. Instead, $z\sim 0.1-1.5$ variability-selected AGNs lie in between the $M_{\rm{BH}}-M_{\ast}$ relation for local inactive early-type galaxies and local active galaxies. This result agrees with most previous studies of $M_{\rm{BH}}-M_{\ast}$ relation for AGNs at similar redshifts, regardless of selection technique. We demonstrate that studies of variability selected AGN provide critical insights into the low-mass end of the $M_{\rm{BH}}-M_{\ast}$ relation, shedding light on the occupation fraction of that provides constraints on early BH seeding mechanisms and self-regulated feedback processes during their growth and co-evolution with their hosts.
\end{abstract}



\section{Introduction} \label{sec:intro}

Local scaling relations between supermassive black hole (SMBH) mass $M_{\rm{BH}}$ and their host galaxy properties (e.g., total galaxy stellar mass, bulge stellar mass, stellar velocity dispersion, bulge luminosity: $M_{\rm{BH}}-M_{\ast}$, $M_{\rm{BH}}-M_{\ast,\rm{bulge}}$, $M_{\rm{BH}}-\sigma_{\ast}$, $M_{\rm{BH}}-L_{\ast,\rm{bulge}}$) in both active and inactive galaxies are key to our understanding of SMBH-host galaxy coevolution (e.g., \citealt{Magorrian1998,Haehnelt1998,Kormendy2013,Reines2015}). These empirical scaling relations are widely taken as evidence for some form of self-regulated feedback between black hole accretion and star formation in active galactic nuclei (AGNs). However, there is no consensus model of how AGN feedback operates (e.g., \citealt{Fabian2012}). Observational measures of the SMBH--host scaling relations over a wide range of redshifts and luminosities are critical for placing new constraints on such models (e.g., \citealt{Ricarte2018}). In addition to informing theoretical models of AGN feedback, the low-mass end of the SMBH-host scaling relations may be sensitive to the initial mass function of high redshift SMBH seeds and their subsequent growth modes \citep{Volonteri2009,Natarajan2011}.

While the existence of an $M_{\rm{BH}}-\sigma_{\ast}$ relation may be universal feature of structure formation (e.g., \citealt{vandenBosch2016}), the slope and normalization of the local $M_{\rm{BH}}-M_{\ast}$ scaling relations differ significantly for actively accreting AGNs and galaxies hosting inactive black holes \citep{Reines2015}. The SMBH-host galaxy scaling relations have been studied in AGNs selected over a range of wavelengths and redshifts (e.g., \citealt{Lilly2007,Merloni2010,Civano2016,Suh2020,Ding2020,Li2021,Zhuang2023,Mezcua2023,Mezcua+2024, Li2023}). In our previous work, we studied $z<4$ variability-selected AGNs from the HSC-SSP program \citep{Kimura2020} in the COSMOS field and their $M_{\rm{BH}}-M_{\ast}$ scaling relation \citep{Burke2024hsc}. These results generally confirm that these intermediate and higher redshift AGNs have over-massive black holes compared to the local ($z<0.055$) AGN relation of $M_{\rm{BH}}/M_{\ast} \sim 0.025\%$ \citep{Reines2015}, regardless of the selection method. Instead, non-local AGNs appear to follow the $M_{\rm{BH}}-M_{\ast}$ relation more closely for local \emph{inactive} galaxies.


Meanwhile, \emph{James Webb Space Telescope} ({\emph{JWST}}) $z \sim 4-7$ AGNs have black hole masses $\sim 10 - 100$ times more massive compared to local AGNs with comparable stellar mass hosts (\citealt{Harikane2023,Maiolino2023,Kocevski2023,Ubler2023}). Recently, \citet{Kokorev2023} identified a $z=8.5$ AGN with a $M_{\rm{BH}}/M_{\ast}$ ratio of at least $\sim 30$ percent, and \citet{Bogdan2023,Natarajan2023,Goulding2023} report a $z\approx10.1$ quasar UHZ1 and the source GHZ9 at $z\approx10.4$ \citep{Kovacs+2024}, both with $M_{\rm{BH}}/M_{\ast} \sim 1$. The $z=10.6$ source GN-z11 also lies above the $M_{\rm{BH}}-M_{\ast}$ for local AGNs \citep{Maiolino2023}. Elevated $M_{\rm{BH}}/M_{\ast}$ ratios at redshifts of $9<z<12$ are a predicted outcome of heavy black hole seed formation in the early Universe via direct collapse of pristine gas \citep{Natarajan+2017}. On the other hand, \citet{Li2025} identified a population of $z\sim 3-5$ AGNs with $M_{\rm{BH}}/M_{\ast}$ ratios that are more in-line with the relation for local inactive galaxies than these earlier JWST results (but still more massive than the \citet{Reines2015} relation for local AGNs), which is consistent to the AGN sample of \citet{Sun2024} selected from JADES result at a lower redshift of $1\lesssim z\lesssim4$.

An important caveat is that these observations are strongly affected by selection biases. For example, AGNs selected by luminosity can produce a false redshift evolution in the host galaxy scaling relations \citep{Lauer2007}. Statistical modeling by \cite{Li+2024} shows how a combination of selection biases and measurement uncertainties can result in elevated $M_{\rm{BH}}/M_{\ast}$ ratios for $z > 6$ {\emph{JWST}} AGNs. Furthermore, uncertainties from single-epoch virial black hole mass measurements can lead to the systematic overestimation of black hole masses, especially at the high mass end of these relations \citep{Shen2010}. Therefore, it is imperative to select AGNs with lower masses and luminosities over a wide range of redshifts to mitigate these selection biases (e.g., \citealt{Izumi2019,Izumi2021,Suh2020}). 

Optical variability is uniquely suited to uncover faint, low-mass AGNs in dwarf galaxies, which are otherwise missed by traditional selection methods \citep{Villforth2012,Baldassare2018,Baldassare2020,Halevi2019,Guo2020,Burke2020tess,Burke2022des,Burke2023,Ward2022}. In this paper, we obtain black hole masses, redshifts, and robust stellar masses for variability-selected low-mass AGNs from \citet{Burke2022des}. These sources were identified from light curves from the Dark Energy Survey supernova program (DES-SN; \citealt{Vincenzi2024}), a subset of the wide-field DES \citep{DES2016}. About 31 percent of the sources are detected in the X-ray. Using these data, we measure the $M_{\rm{BH}}-M_{\ast}$ relation at $z\sim 0.1-1.5$. Our sample has a bolometric luminosity range of $L_{\rm{bol}} \sim 10^{44-46}$ erg s$^{-1}$, comparable to the COSMOS X-ray selected sample of \citet{Suh2020}. In combination with previous work \citep{Burke2024hsc}, we find that this sample of $z\sim 0.1-3.5$ variability-selected AGNs have overmassive black holes compared to the local AGN relation of \citet{Reines2015}, broadly consistent with most previous studies of AGNs at similar median redshifts ($z\sim 0.5-1$) selected with other techniques \citep{Merloni2010,Mezcua2023,Li2021,Zhang2023,Zhuang2023,Stone2023,Mountrichas2023,Tanaka2024}. 

This paper is organized as follows. In \S\ref{sec:data}, we describe our procedure for collating archival spectra and photometry and constructing a spectroscopic redshift database for our variable AGN sample. In \S\ref{sec:sed}, we describe our procedure for estimating the stellar masses of the host galaxies by fitting the spectral energy distributions (SEDs) to the broad-band photometry of the host galaxy in the presence of an AGN. In \S\ref{sec:spec}, we describe our black hole mass estimates derived from fitting the broad lines in the spectra. In \S\ref{sec:relation}, we place these sources on the $M_{\rm{BH}}-M_{\ast}$ relation and compare our results with previous work. We discuss our results \S\ref{sec:discussion} and their implications in \S\ref{sec:conclusion}.

\subsection{Dark Energy Survey Deep Field Data}

In this work, we use the sample of variable AGNs selected by \citet{Burke2022des} from $g$-band light curves from the DES-SN program. Both resolved and unresolved sources were selected, using difference imaging for resolved sources that tend to be at lower redshifts\footnote{These light curves were re-computed using a single template for all observing seasons, which differs from the DES-SN difference imaging light curves used for the transient search pipeline \citep{Kessler2015}.}. DES-SN observed 10 fields over 6 years with a cadence of about 7 days (with seasonal gaps) in the $griz$ bands during the observing. The DES-SN observations generally took place when the seeing was poor. The single-epoch photometric precision is $g \sim 24.5$. These fields have been continuously observed since then \citep{Zhuang2024helm} and will be a part of the upcoming Legacy Survey of Space and Time (LSST) using the Vera C. Rubin Observatory \citep{Ivezic2019}. Our parent sample is restricted to 3 DES-SN fields with 8-band ($ugrizJHK_S$) deblended, stacked model-based photometry \citep{Hartley2022} to a uniform depth of $i = 25$ that are critical to obtaining stellar mass estimates. These fields overlap with the Chandra Deep Field-South \citep{Luo2017}, the XMM Large Scale Structure survey fields \citep{Garcet2007}, and the European Large Area ISO Survey \citep{Oliver2000}. The deep fields of our parent sample of AGNs from \citet{Burke2022des} are denoted C3, X3, and E2, respectively. The total C3+X3+E2 area with optical+NIR photometry is $\sim 4.6$ deg$^2$.

\section{Data Analysis} \label{sec:data}

\subsection{Spectra Database} 

\begin{table}
\scriptsize
\centering
\caption{List of sources of the optical/NIR spectroscopic redshifts for the DES-SN AGNs. \label{tab:redshifts}}
\begin{tabular*}{\linewidth}{cll}
\hline\hline
Count & Reference & Name \\
\hline
50 & \citet{Menzel2016} & SDSS \& WISE AGN \\
44 & \citet{Burke2022des} & OzDES$^a$ \\
28 & \citet{Tie2017} & WISE \& DES AGN \\
16 & \citet{Mao2012} & ATLAS \\
13 & \citet{Silverman2010} & CDF-S Faint X-ray \\
12 & \citet{Ahumada2020} & SDSS DR16 \\
6  & \citet{Alam2015} & SDSS-III DR12 \\
6  & \citet{Sacchi2009} & Spitzer/SWIRE \\
5  & \citet{Baldry2018} & GAMA DR3 \\
3  & \citet{Zou2023} & XMM-SERVS \\
3  & \citet{Paris2018} & SDSS DR14Q \\
3  & \citet{Garilli2014} & VIPERS PDR-1 \\
3  & \citet{Burke2022des} & PanS$^a$ \\
3  & \citet{Burke2022des} & ACES$^a$ \\
2  & \citet{Cheng2021} & Low-$z$ faint galaxies \\
2  & \citet{Scodeggio2018} & VIPERS PDR-2 \\
2  & \citet{Suh2015} & FMOS AGN \\
2  & \citet{Cooper2012} & ACES \\
2  & \citet{Eales2009} & BLAST \\
2  & \citet{Burke2022des} & SDSS$^a$ \\
1  & \citet{Lyke2020} & SDSS DR16Q v4 \\
1  & \citet{Lacy2013} & Spitzer AGN \\
1  & \citet{LeFevre2013} & VVDS \\
1  & \hspace{-0.12\linewidth}\makecell[{{p{0.35\linewidth}}}]{\citet{Hernan-Caballero2011}} & Spitzer \\
1  & \citet{Xia2011} & Magellan LDSS-3 \\
1  & \citet{Burke2022des} & VVDS$^a$ \\
1  & \citet{Burke2022des} & SNLS$^a$ \\
1  & \citet{Burke2022des} & GAMA$^a$ \\
1  & \citet{Burke2022des} & 2dFG$^a$ \\
\hline
\end{tabular*}
\footnotesize{\textsc{Note.} --- ($a$) Data remain unchanged from \citet{Burke2022des}.}
\end{table}

\begin{figure}
\centering
\includegraphics[width=0.49\textwidth]{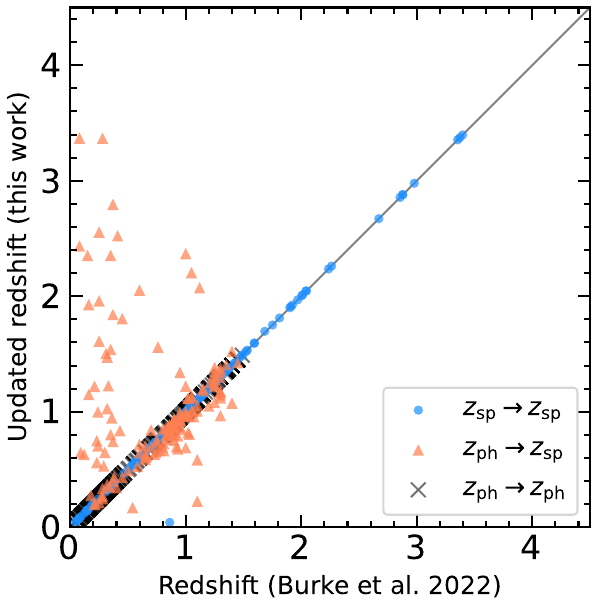}
\caption{Updated redshifts in this work from SIMBAD, OzDES, and DESI v.s. original redshifts from \citet{Burke2022des}. The redshifts updated from photometric to spectroscopic are labeled using orange triangle symbols, and the unchanged or updated photometric and spectroscopic redshifts are shown as gray crosses and blue dots, respectively. A number of previous catastrophically underestimated photometric redshifts have been updated \label{fig:zupdated}}
\end{figure}

\begin{figure}
\centering
\includegraphics[width=0.49\textwidth]{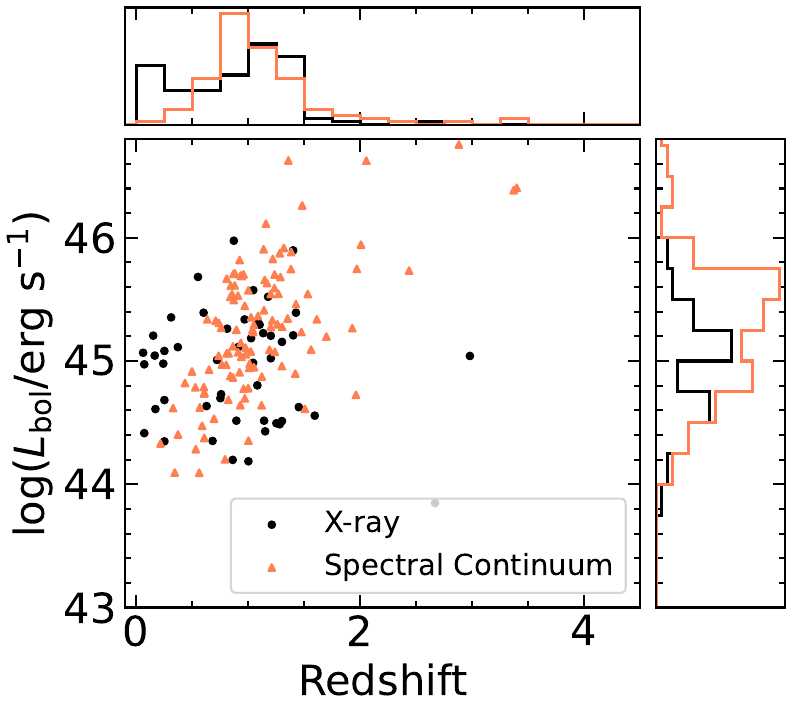}
\caption{The bolometric AGN luminosity, estimated as ten times the 2–10 keV X-ray luminosity (following \citealt{Duras2020}), is plotted against our updated redshifts. The distributions of redshift and bolometric luminosity are displayed in the upper and right panels, respectively.} \label{fig:Lbolz}
\end{figure}

\begin{figure}
\centering
\includegraphics[width=0.49\textwidth]{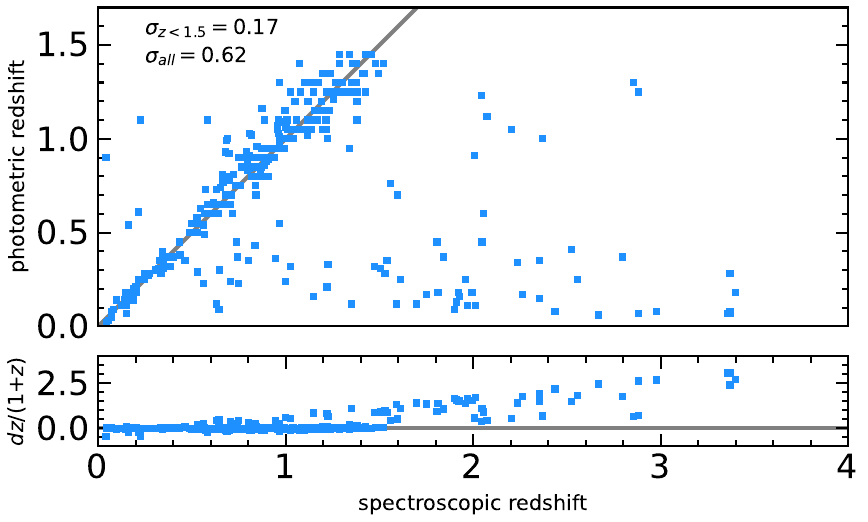}
\caption{Comparison of the photometric redshifts from \citet{Burke2022des} and the spectroscopic redshifts obtained in this work. The photometric redshifts are determined using the approach outlined in \citet{Yang2017}. The root mean square error (RMSE) values of all the sources and $z_{\rm{spec}}<1.5$ sources are displayed in the upper left corner of the figure panel. A gray $y=x$ line is included for reference.\label{fig:photoz}}
\end{figure}

\citet{Burke2022des} and \citet{Hartley2022} compiled spectroscopic redshifts in the DES deep fields, which we supplement here using the SIMBAD database \citep{Wenger2000} and the OzDES second data release. We identified 99 additional spectroscopic redshifts from the literature that were indexed at the time of writing by SIMBAD using an \textsc{astroquery} \citep{Ginsburg2019} SIMBAD search (see references in Table~\ref{tab:redshifts}). We obtained additional from DESI and OzDES public catalogs. We restricted our SIMBAD search to good quality spectroscopic redshifts from optical or near-infrared (NIR) spectra using the flags ${\rm RVZ\_WAVELENGTH} == {\rm 'O'}$ or ${\rm RVZ\_WAVELENGTH} == {\rm 'N'}$ and ${\rm rvz\_qual} != {\rm'E'}$. The first constraint restricts the search to redshifts from optical/NIR spectra. The second constraint excludes photometric redshifts. Out of 706 AGN candidates, we removed 10 spectroscopically-classified variable stars (ID=14, 84, 327, 328, 426, 544, 626, 673, 669, and 680) from our parent sample using the OTYPE key. ID=426 is a spectroscopically-confirmed RR Lyrae variable star, and ID=669 is a spectroscopically-confirmed cataclysmic variable star. 
Finally, we downloaded the publicly available spectra from these sources (see Appendix~\ref{sec:availibility}). We also obtained spectroscopic redshifts from the Dark Energy Spectroscopic Instrument (DESI) Early Data Release \citep{DESI2023}, which have not yet been indexed by SIMBAD at the time of writing. We assign fiducial spectroscopic redshifts according to the following priority: DESI (5 sources) $>$ OzDES (147 sources) $>$ SIMBAD (99 sources). For example, when both DESI and SIMBAD have redshifts, we adopt DESI's one. 

If no spectroscopic redshift was found, we use the photometric redshifts from \citet{Burke2022des} (407 sources; \S\ref{sec:photoz}). Our updated redshifts are shown in Figure~\ref{fig:zupdated}. The bolometric luminosity and redshift distributions are shown in Figure~\ref{fig:Lbolz}. A table showing the sources of the public spectroscopic redshifts for the DES deep field AGNs is shown in Table~\ref{tab:redshifts}. We found no inconsistent spectroscopic redshifts between our adopted spectroscopic redshifts in this work and the spectroscopic redshifts of \citet{Burke2022des}. To include cases where a single source has more than one available spectrum from different programs, we always repeat the matching between the DES deep field AGNs and the spectroscopic sample when downloading the spectra from publicly available sources.

\subsection{Photometric redshifts} \label{sec:photoz}
For sources without a reliable spectroscopic redshift, we use the photometric redshifts from \citet{Burke2022des} derived from the Skew-t method of \citet{Yang2017}. This approach uses the asymmetries in the relative flux distributions as a function of redshift and magnitude and was trained on a sample of AGNs and galaxies. Using our newly compiled spectroscopic redshifts as our ground truth, we estimate the scatter in the photometric redshifts following \citet{Burke2024hsc} as $\sigma = 1.4826$ MAD, where MAD is the median absolute deviation, which is robust to outliers. We find $\sigma = 0.62$ for the entire dataset, but $\sigma = 0.17$ at spectroscopic redshifts of $< 1.5$. This indicates a large outlier fraction due to contamination from high-redshift quasars. A comparison between the photometric redshifts and our updated spectroscopic redshifts for the DES deep field variable AGNs are shown in Figure~\ref{fig:photoz}. 

\section{SED fitting}  \label{sec:sed}

We recomputed the stellar masses for the \citet{Burke2022des} DES deep field AGNs using our newly compiled spectroscopic redshifts. We use the $ugrizJHK_S$ de-blended, stacked model-based photometry from \citet{Hartley2022}, which is publicly available. The photometry is derived from DES imaging along with supplemental $u$-band DECam imaging that was obtained in these fields. The $JHK_S$ imaging of E2, C3, and X3 are from the VIDEO survey \citep{Jarvis2013}. We also extended the photometry to UV and IR by matching the \citet{Burke2022des} DES deep field AGNs to the Revised Catalog of GALEX UV Sources \citep{bianchi2017} and Spitzer Enhanced Imaging Products Source (SEIP) List \citep{slphotdr4}, respectively. For SEIP, we use all matched Spitzer data and associated ALLWISE data with quality flags of either A, B, or C. In addition, the X-ray 2$-$7 keV photometry is from version 2.0 of the \emph{Chandra} Source Catalog \citep{evans2010} and the X-ray 0.2$-$12 keV photometry is from XMM-Newton serendipitous survey \citep{webb2020}. We converted X-ray fluxes from $\text{erg\ s}^{-1}\ \text{cm}^{-2}$ to $\text{mJy}$ using the method provided by \citet{Yang2020}. All catalogs are matched within 1 arcsec. The inclusion of the optical to mid-infrared photometry is essential to constrain the star formation and reprocessed dust emission. Additionally, including the X-ray photometry helps constrain the contribution from AGN emission \citep{Yang2020}. This multiwavelength SED helps to eliminate \emph{some} degeneracies between star formation and AGN emission, which can lead to spurious stellar mass estimates (e.g., \citealt{Burke2024hsc}). 

We use version 2022.1 of the \textsc{cigale} code \citep{Burgarella2005,Noll2009,Boquien2019,Yang2020,Yang2022} for SED fitting. This version includes X-ray fitting modules from \textsc{x-cigale} and incorporates detailed AGN emission models that have been thoroughly tested on both galaxies and AGNs \citep{Yang2020,Yang2022}. The codes enforce a self-consistent energy balance between different emission and absorption processes across the electromagnetic spectrum. A large set of models is involved and fitted to the data, enabling the estimation of stellar mass, star formation rate (SFR), and AGN contribution through a Bayesian-like analysis of the likelihood distribution.

We adopt a delayed exponential star formation history, varying the $e$-folding time and the age of the stellar population while assuming solar metallicity. However, systematic uncertainties of approximately 0.3 dex can arise from different choices of initial mass function, stellar population models, and star formation histories \citep{Conroy2013}. \citet{Zou2022} found that variations in \emph{parametric} star formation histories lead to systematic stellar mass differences of about 0.1 dex for AGNs in the redshift range $z = 0 - 6$. 

For our analysis, we use the widely adopted \citet{Chabrier2003} initial stellar mass function, along with the stellar population models from \citet{Bruzual2003} and the nebular emission template of \citet{Inoue2011}. Dust extinction is modeled using the \citet{Leitherer2002} extension of the \citet{Calzetti2000} attenuation law, while dust emission follows the \citet{Draine2014} updates to the \citet{Draine2007} model. For AGN emission, we apply the SKIRTOR clumpy two-phase torus model \citep{Stalevski2012, Stalevski2016}, incorporating additional polar extinction. 

We assume a Type-1-like inclination angle of $30^\circ$ for our optically variable AGNs. This choice is supported by prior studies, which indicate that different viewing angles are largely degenerate with the typical values of 30 and 70 degrees for Type 1 and Type 2 AGNs, respectively \citep[e.g.,][]{Mountrichas2021, Padilla2022}. Table 2 of \citet{Burke2024hsc} lists the \textsc{cigale} input parameters, with example SED fitting results presented in Figure~\ref{fig:seds}. The derived stellar masses range from $M_{\ast} \sim 10^{7-11.5}\ M_{\odot}$.

\subsection{Reliability of stellar mass estimates}
\label{sec:Mstarrecovery}

\begin{figure*}
\centering
\includegraphics[width=0.48\textwidth]{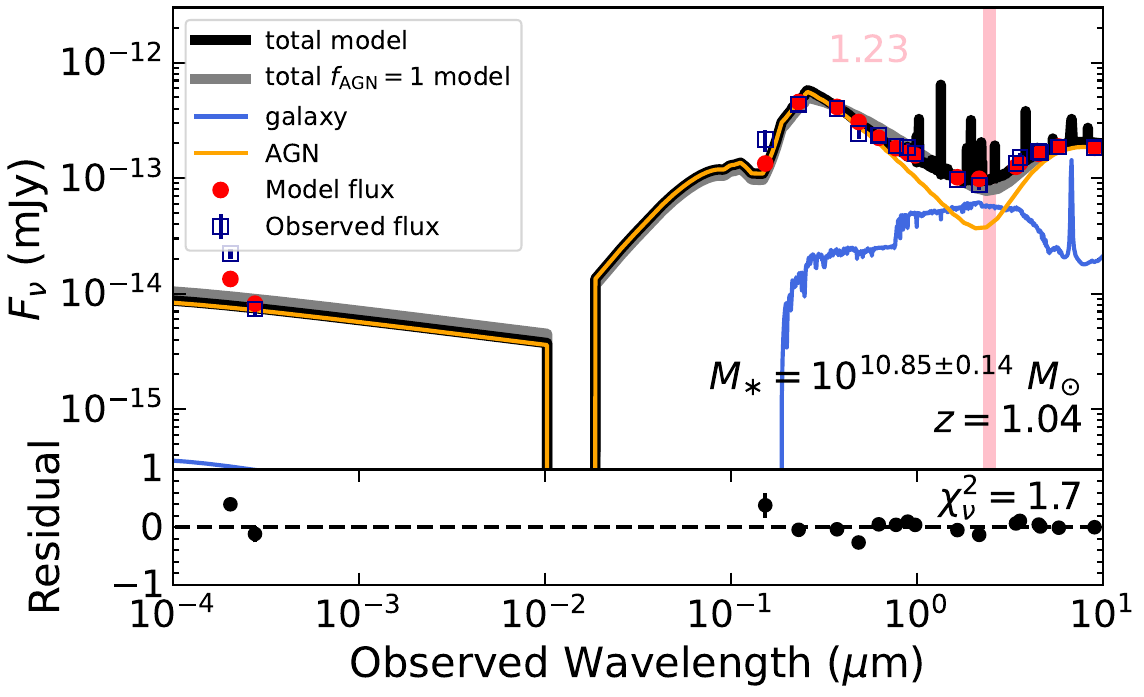}
\includegraphics[width=0.48\textwidth]{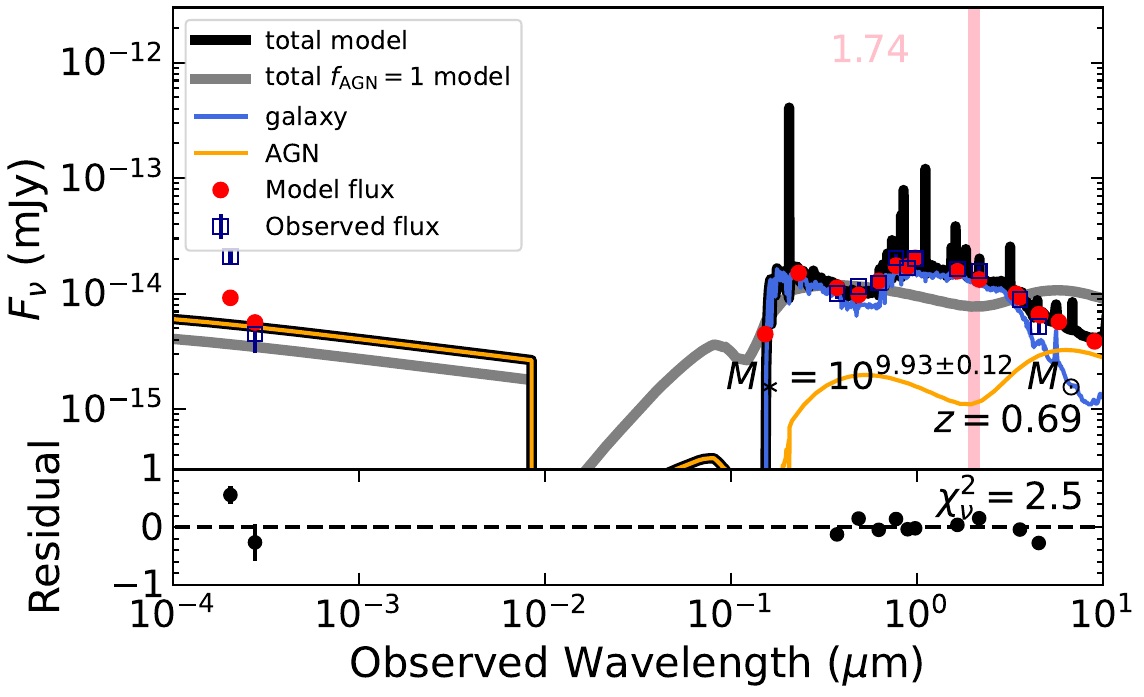}
\caption{\emph{Left panel:} An example SED fitting result for an AGN-dominated source (\citet{Burke2022des} ID = 43). In this case, including a stellar emission component does not significantly improve the fit, meaning that the stellar emission and related parameters, such as stellar mass and star formation history, cannot be reliably constrained. \emph{Right panel:} An example SED fitting result for a source that is not AGN-dominated (\citet{Burke2022des} ID = 83). Here, adding a stellar emission component improves the best-fit $\chi_\nu^2$, and the resulting best-fit stellar emission shows a significant excess over the AGN continuum emission near $\sim 1.2$ $\mu$m, allowing for a reliable stellar mass estimate. The uncertainties on the stellar mass are statistical only, as provided by \textsc{cigale}. The vertical red line near $1.2$ $\mu$m marks the region where AGN emission is at a minimum. The red number represents the ratio of star formation (SF) emission (blue) in the SF$+$AGN model to the total emission (black) in the AGN-dominated model (gray). Both redshifts are spectroscopic. \label{fig:seds}}
\end{figure*}


\begin{table*}
\caption{Properties of DES-SN variable AGNs from SED and spectral fitting analysis.   \label{tab:mass}}
\center
\begin{tabular}{cccccccccccccc}
\hline\hline
ID & RA & DEC & g-mag & $z_{{\rm best}}$ & $z_{{\rm ph}}$ & $\log L$ & $\log M_{\rm{BH}}$ & $\log M_{\rm{BH, err}}$ & $\log M_{\ast}$ & $\log M_{\ast, {\rm{err}}}$ & ${\rm{SF}}_{\rm{ex}}$ & $\chi^2_\nu$ & class \\
(1) & (2) & (3) & (4) & (5) & (6) & (7) & (8) & (9) & (10) & (11) & (12) \\
 & deg & deg & AB mag &   &   & $\log$ erg s$^{-1}$ & $\log M_{\odot}$ & $\log M_{\odot}$ & $\log M_{\odot}$ & $\log M_{\odot}$ & \\
\hline
1 & 52.1408 & -27.1823 & 23.7 & 1.02 & 1.02 &  &  &  & 10.96 & 0.06 & 1.4 & 4.8 &  \\
2 & 52.4633 & -27.1826 & 23.4 & 0.24 & 0.24 &  &  &  & 8.63 & 0.24 & 1.0 & 1.6 &  \\
3 & 52.436 & -27.1854 & 23.4 & 1.09 & 1.09 &  &  &  & 10.39 & 0.12 & 0.9 & 3.0 &  \\
4 & 52.1894 & -27.1881 & 21.5 & 1.3 & 1.3 &  &  &  & 10.16 & 0.59 & 1.3 & 6.3 &  \\
\vdots & & & & & \\
706 & 36.3473 & -5.0628 & 19.8 & 0.1909 & 0.2 &  &  &  & 10.47 & 0.11 & 9.6 & 6.6 &  \\
\hline
\end{tabular}
\begin{flushleft}
{\sc Note.} Column (1): Identifier from Table 4 of \citet{Burke2022des}. Column (2): RA. Column (3): Dec. Column (4): $g$ band AB magnitude. Column (5): Best redshift. Column (6): Photometric redshift. Column (7): Bolometric luminosity from the spectral continuum. Column (8): Virial black hole mass. Column (9): Inferred \textsc{cigale} stellar mass. Column (10): Excess SF over an AGN dominated model (if $>1.2$, stellar masses are considered reliable). Column (11): Best-fit reduced \textsc{cigale} model $\chi^2$ (recommend $<5$). Column (12): Visual classification of spectrum (see \S\ref{sec:vis}). All uncertainties are $1\sigma$ statistical errors from fitting. This table is published in its entirety in the published version. Only a portion is shown here.
\end{flushleft}
\end{table*}

For galaxies where AGN emission is significant, constraining or subtracting this emission is essential for accurately modeling the star formation and deriving reliable stellar masses. This process is particularly challenging, and a robust stellar mass estimation often requires high-resolution optical/NIR imaging combined with source profile fitting to remove the AGN point source from the underlying host galaxy. Alternatively, an AGN template can be fit to the spectral continuum and emission lines, then scaled to the photometry to model the AGN contribution \citep{Reines2015}. When high-resolution imaging or spectra are unavailable, SED fitting using broad-band catalog photometry can be applied with certain limitations.

The UV/optical emission from an unobscured AGN accretion disk can be highly degenerate with stellar emission. In cases where the SED is dominated by an unobscured AGN, the stellar emission component becomes overwhelmed, making it difficult to constrain stellar properties such as stellar mass and star formation history reliably \citep[e.g.,][]{Merloni2010,Ciesla2015}. To assess whether stellar emission, and thus stellar mass, can be constrained from SED fitting, we employ a model comparison method following \citet{Burke2024hsc}.

This approach involves first fitting the SED using an AGN-dominated model, setting the AGN fraction to $f_{\rm{AGN}}=0.9999$, where $f_{\rm{AGN}}$ represents the AGN contribution in the observed-frame $0.5-1$ $\mu$m range. This ensures that the SED is fully dominated by AGN emission, as AGN contribution is minimal at these wavelengths. Due to technical constraints in \textsc{cigale} \citep{Yang2020}, $f_{\rm{AGN}}$ cannot be set exactly to unity, so we round it to 1 for clarity in the rest of this paper. Next, we re-fit the SED using a mixed AGN+stellar emission model, allowing $f_{\rm{AGN}}$ to vary between 0 and 1 in the observed-frame $0.1-0.3$ $\mu$m range, where most of our spectra exhibit significant AGN continuum emission. Since our sources are identified based on optical light curves, AGN emission is expected to be prominent in this range. Indeed most of our spectra show significant AGN continuum emission at these wavelengths because the sources have been identified from optical band light curves.  

We consider stellar masses to be reliable only for SEDs that show significant stellar emission at rest-frame 1.2 $\mu$m, which cannot be reproduced by a fully AGN-dominated model. To quantify this, we compute the ratio of the total model fit from the AGN+stellar model to that of the AGN-dominated model. A stellar mass is deemed reliable if this ratio exceeds 1.2 \citep{Burke2024hsc}. This method is similar to some previous approaches \citep{Merloni2010,Suh2020,Burke2022des} and effectively removes SEDs that are degenerate with AGN emission at rest-frame 1.2 $\mu$m, where AGN contamination is minimized.

Our approach improves upon previous stellar mass estimates in \citet{Burke2022des} by incorporating X-ray data, which significantly constrains the AGN contribution to the SED \citep{Yang2022,Burke2022des}. Additionally, comparing reduced $\chi^2$ values alone, as done in \citet{Burke2022des}, is less reliable because $\chi^2$ depends on photometric uncertainties, and model parameter choices (e.g., over-fitting stellar emission), and does not always indicate a better fit in the optical-NIR region, where stellar emission is strongest and stellar masses are derived \citep{Burke2022des}. A detailed comparison between the two approaches is presented in Appendix~\ref{sec:dchi2}.

Figure~\ref{fig:seds} presents example SEDs with both reliable and unreliable stellar mass estimates. The scatter in the derived stellar masses is typically around 0.2 dex, an uncertainty not fully accounted for in \textsc{cigale}, which tends to underestimate stellar mass uncertainties. To compensate, we add 0.2 dex in quadrature to the \textsc{cigale} uncertainties in all figures throughout this paper. We adopt the ``Bayesian-like'' parameter estimates and uncertainties from \textsc{cigale}, which are derived by weighting each model solution by $\exp(-\chi^2/2)$ \citep{Boquien2019}, with final values obtained from the likelihood-weighted mean and standard deviation.

\subsection{Detection limits}

\begin{figure*}
\centering
\includegraphics[width=0.95\textwidth]{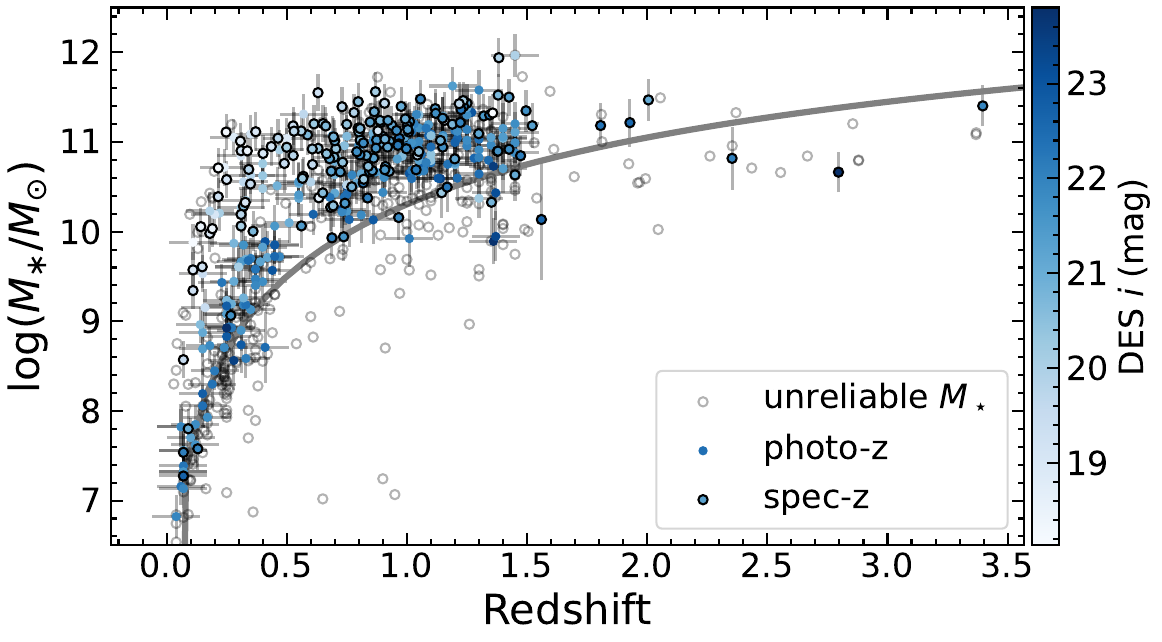}
\caption{Host galaxy stellar mass versus redshift for DES-SN variable AGNs with stellar mass estimates from broad-band SED fitting using \textsc{cigale}. gray circles: unreliable stellar mass ($\chi_\nu^2>5$ or SF excess $<1.2$). Colored circle symbols with black border: reliable stellar mass calculated from spectroscopic redshift. Colored circle symbols without black border: reliable stellar mass calculated from the photometric redshift. Gray Curve: theoretical detection limit. \label{fig:massredshift}}
\end{figure*}

Figure~\ref{fig:massredshift} presents our derived stellar mass estimates versus redshift for the DES-SN AGNs. The gray curves represent the theoretically predicted stellar mass detection limits, following \citet{Burke2023}. These stellar mass horizon curves are calculated by assuming a limiting detectable variability amplitude of 0.1 magnitudes and incorporating the modified correlations between optical variability amplitude and mass \citep[e.g.,][]{MacLeod2010}, as described in \citet{Burke2023}. The predicted detection limits are based on a typical variability amplitude of 0.1 mag and a photometric precision from \citet{Ivezic2019}, with a DES single-epoch limiting magnitude of $g=24.5$ \citep{Burke2022des}.

\section{Spectral Measurements} \label{sec:spec}

\subsection{Publicly available spectroscopic data}

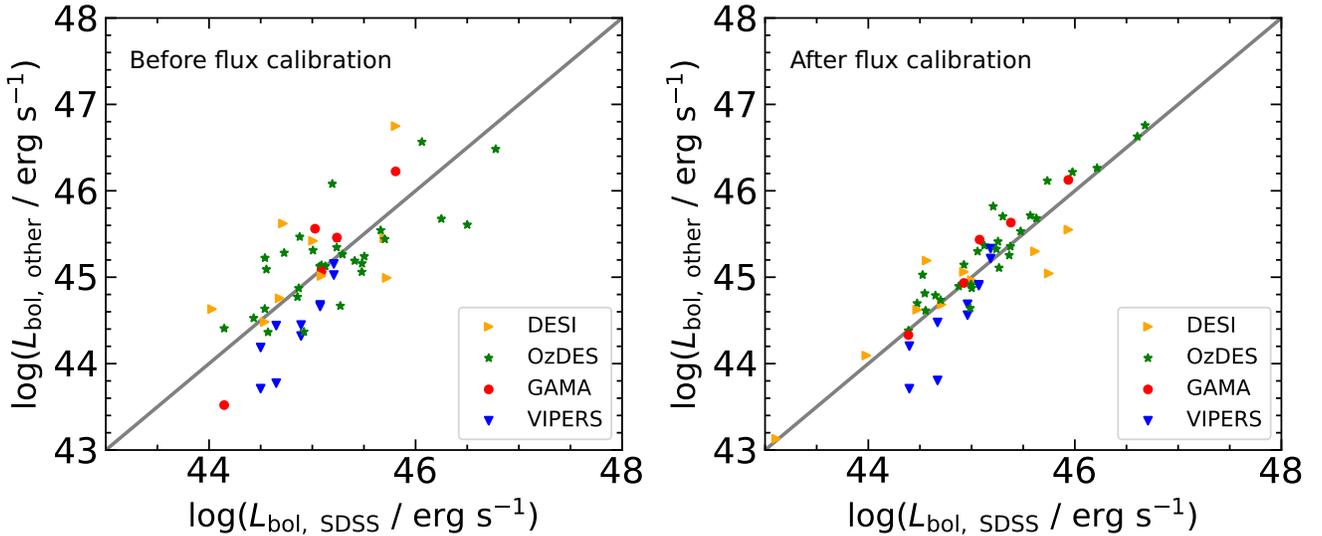
\begin{figure*}
\centering
\includegraphics[width=0.48\textwidth]{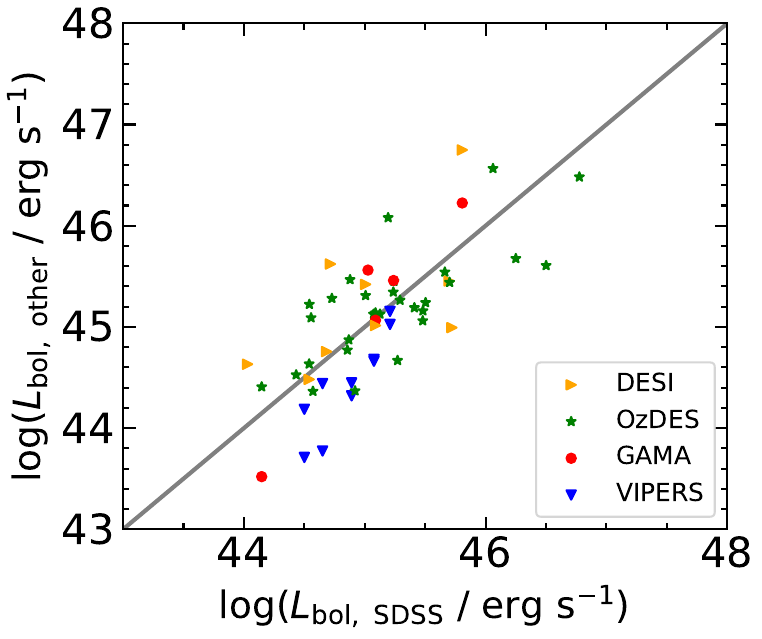}
\includegraphics[width=0.48\textwidth]{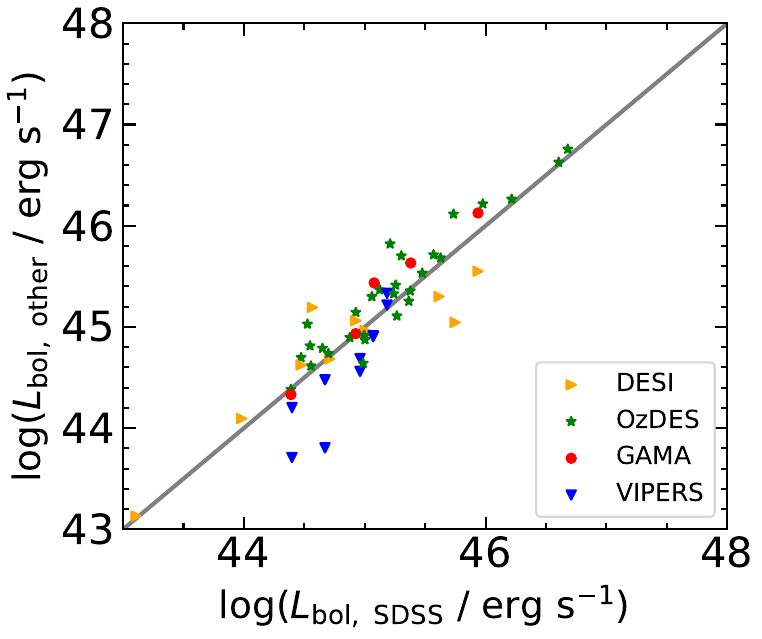}
\begin{picture}(0,0) 
    \put(-450,180){{\usefont{T1}{DejaVuSans-TLF}{m}{n}Before flux calibration}} 
    \put(-200,180){{\usefont{T1}{DejaVuSans-TLF}{m}{n}After flux calibration}} 
\end{picture}
\caption{Demonstration of absolute flux calibration of our spectra. Each data point is a single DES deep field AGN with one spectrum from SDSS and and at least one from another source. The bolometric luminosity from SDSS spectral continua is plotted against the bolometric luminosity from other spectra (see Table~\ref{tab:redshifts} for data source list) before (\emph{left}) and after (\emph{right}) performing absolute flux calibration to the \citet{Hartley2022} photometry. Although AGN variability can contribute to some of the scatter across surveys, our resulting flux-calibrated spectra appear to be more consistent as evidenced by the significantly reduced scatter. The right panel only includes those with successful flux calibration. \label{fig:fluxing}}
\end{figure*}

We obtained publicly available spectra for our parent sample of DES deep-field AGNs from Table~\ref{tab:redshifts}. The list of downloaded spectra differs slightly from our spectroscopic redshift database in Table~\ref{tab:redshifts}, as not all spectroscopic data are publicly accessible. The spectra were collected from the following programs: the extended CDF-S faint X-ray source survey, GAMA, OzDES, VIPERS, SDSS, and DESI. Appendix~\ref{sec:availibility} provides links for downloading these spectra. A brief description of each dataset is given below.


The extended CDF-S faint X-ray source survey obtained spectroscopic redshifts of 283 Chandra sources up to $z\sim4$. This program used the VLT/VIMOS and Keck/DEIMOS instruments. The VIMOS spectra cover a wavelength range of $\sim 5500-9600$ \AA\ with a spectral resolution of $R \sim 580$ (red grism) and $\sim 3700-6700$ \AA\ with $R \sim 180$ (blue grism). The DEIMOS spectra cover $\sim 4600-9700$ \AA\ and have a spectral resolution of $3.5$ \AA\ ($\approx 125$ km s$^{-1}$), which is sufficient for measurements of the width of the broad lines \citep{Silverman2010}.

OzDES is a spectroscopic survey aiming to obtain the redshifts of DES sources using the 4m Anglo-Australian Telescope with a Two Degree Field (2dF) multi-object fiber positioner. This instrument covers $\sim 3800-8900$ \AA\ with a spectral resolution of $R\sim1800$ \citep{Childress2017}.

GAMA is a large spectroscopic survey of low-redshift galaxies. It covers galaxies of $z\lesssim0.8$ with $r < 19.8$ mag. Its spectra cover a wavelength range of $\sim 3470 - 8850$ \AA. Its resolution ranges from $R\sim1000$ at the blue end to $R\sim1600$ at the red end, with a pixel size of $1.04\AA$ \citep{Liske2015}.

VIPERS is a public extragalactic redshift survey using the VLT/VIMOS instrument. This project obtained the photometry, spectra, and redshifts of galaxies bounded by $i_{AB}<22.5$ with a redshift range $0.5<z<1.2$. Its spectra cover $\sim 5500 - 9500$ \AA\ with a resolution of $R\sim220$ \citep{Scodeggio2018}.

SDSS is a wide-field spectroscopic survey, covering galaxies and quasars using various selection methods depending on the specific SDSS survey subset. Its spectroscopy spans approximately 3600 to 9800 Å, with a spectral resolution of $R \sim 2000$ \citep{Ahn2012, Almeida2023}.

DESI is another ongoing wide-field spectroscopic survey that targets galaxies, quasars, and Milky Way stars. The instrument provides spectral coverage from approximately 3600 to 9800 Å with a resolution of $R \sim 2000-5500$ \citep{DESI2023}.

\subsection{Flux calibration and data cleaning}

In total, we have 1040 spectra with varying quality and wavelength coverage. Below, we outline our procedure for calibrating and cleaning the data. A reliable estimate of flux uncertainties is crucial for least squares minimization and $\chi^2$ estimation \citep{Burke2024hsc}. When an error spectrum is not provided, we approximate uncertainties using the median absolute deviation of the flux spectrum. Wherever possible, we apply data quality masks to account for artifacts and spectral gaps. Each spectrum's fitting results are visually inspected. For sources with multiple spectra, if more than one provides a valid black hole mass estimate, we select the spectrum with the highest signal-to-noise ratio. We deliberately avoid stacking spectra from different surveys for two reasons: (1) combining spectra with different spectral resolutions introduces complications, and (2) lower signal-to-noise spectra often have poorer spectro-photometric calibration, which would propagate through the stack.

Accurate flux calibration is essential for deriving virial black hole masses as these depend on continuum or broad-line luminosities. To minimize systematic differences between spectra from various surveys and instruments, we first apply absolute flux calibration to each spectrum. Specifically, we integrate each spectrum over the available DES+VIDEO $ugrizJHK_S$ bands to generate synthetic photometry, then scale it by the error-weighted mean ratio between synthetic and DES+VIDEO photometry. This method is widely used in the literature (e.g., \citealt{Mallery2012}). Figure~\ref{fig:fluxing} illustrates the improved consistency in spectral calibration across sources with multiple spectra from different surveys and instruments after applying absolute flux calibration with DES+VIDEO photometry.

Other data reduction challenges, such as residual instrumental sensitivity variations, are more difficult to correct. Additionally, we do not account for the effects of variability in the spectra. Based on the scatter observed in Figure~\ref{fig:fluxing}, we estimate a 1$\sigma$ uncertainty of approximately 10 percent in the final absolute flux calibration.

\subsection{Spectral modeling}

We fit the continuum and emission lines in each 1D spectrum using a modified version of the publicly available \textsc{PyQSOFit} code \citep{Guo2018, Shen2019}, which has been widely used to analyze SDSS quasar properties \citep{Shen2011, Wu2022}. However, unlike the SDSS quasar sample, many of our spectra exhibit a significant host galaxy contribution. This is not unexpected, given that our sample consists of fainter quasars that are more host-dominated, whereas the SDSS quasar sample is limited to $i<20$.

To account for this, we first perform quasar/host galaxy decomposition using principal component analysis (PCA) with host galaxy templates from \citet{Bruzual2003}. After subtracting any significant host galaxy component, we model the quasar continuum as a blue power law combined with a third-order polynomial to account for reddening. If including Fe~II emission templates \citep{Vestergaard2001} improves the reduced $\chi^2$ of the continuum fit by at least 20 percent, they are incorporated into the model. The total model consists of the continuum and single or multiple Gaussian components for the emission lines.

Since uncertainties in the continuum model can affect measurements of weak emission lines, we first fit the global continuum in emission-line-free regions. We then fit multiple Gaussian components to the continuum-subtracted spectra in localized regions around the H$\alpha$, H$\beta$, Mg~II, and C~IV emission lines. Given the lower AGN luminosities and reduced continuum signal-to-noise ratios, due to the higher redshifts and faintness of our sources, it is challenging to reliably separate the AGN continuum from the host galaxy contribution. Following \citet{Burke2024hsc}, we use broad-line luminosities rather than AGN continuum luminosities to estimate black hole masses.

\subsection{Deriving Black Hole Masses}

Following \citet{Shen2011}, we estimate black hole masses using the single-epoch virial method \citep[e.g.,][]{Greene2005}, which relies on broad-emission lines. This method assumes that the broad-line region (BLR) is virialized, using the broad-line FWHM as a proxy for the virial velocity and the broad-line or continuum luminosity as an indicator of the BLR radius. Given the spectral quality of some of our sources (Appendix~\ref{sec:conti}), the continuum luminosities are not always well constrained. Therefore, we use broad-line luminosities instead of continuum to estimate black hole masses. Under these assumptions, the black hole mass can be determined using the following equation:

\begin{equation}
\begin{split}
    \log{\left(\frac{M_{\rm{BH}} }{M_{\odot}} \right)} = a + b &\log{ \left( \frac{L_{\rm{br}}}{10^{44}\ \rm{ erg\ s}^{-1}} \right) } \\&+ 2 \log{ \left( \frac{\rm{FWHM}_{\rm{br}}}{\rm{ km\ s}^{-1}} \right) }
\end{split}
\label{eq:BHmass}
\end{equation}
where $L_{\rm{br}}$ and FWHM$_{\rm{br}}$ are the broad-line luminosity and full-width-at-half-maximum (FWHM) with an intrinsic scatter of $\sim0.4$ dex in black hole mass. The coefficients $a$ and $b$ are empirically calibrated against local AGNs with black hole masses measured from reverberation mapping. We adopt the calibrations \citep{Vestergaard2006} from in \citet{Shen2011} derived by \citet{Shaw2012}:
\begin{equation}
    (a, b) = (1.63, 0.49), \quad \rm H\beta
\end{equation}
\vspace{-6mm}
\begin{equation}
    (a, b) = (1.70, 0.63), \quad \rm Mg\ II
\end{equation}
\vspace{-6mm}
\begin{equation}
    (a, b) = (1.52, 0.46), \quad \rm C\ IV.
\end{equation}
For low redshift sources, the broad H$\alpha$ black hole masses are estimated as in \citet{Shen2011} following \citet{Greene2005}:
\begin{equation}
\begin{split}
    \log{\left(\frac{M_{\rm{BH}} }{M_{\odot}} \right)} = 0.379 + 0.43 &\log{ \left( \frac{L_{\rm{br}}}{10^{42}\ \rm{ erg\ s}^{-1}} \right) } \\&+ 2.1 \log{ \left( \frac{\rm{FWHM}_{\rm{br}}}{\rm{ km\ s}^{-1}} \right) }.
\end{split}
\end{equation}

Following \citet{Shen2011}, we determine a fiducial or preferred ``best'' black hole mass based on the ordering described above, considering factors such as wavelength coverage, redshift, and line signal-to-noise ratio (S/N). We include black hole mass estimates only when the broad-line component is detected with $S/N > 2$, as defined in \citet{Burke2023blazar}. If multiple spectra provide valid black hole mass estimates for the same source, we select the spectrum with the highest median per-pixel $S/N$. 

Although an extinction curve is fitted before measuring line luminosities, C~IV and Mg~II-based black hole masses are expected to be more affected by intrinsic reddening compared to H$\beta$-based masses \citep{Shen2019}. Some of these systematic effects may be partially accounted for in the virial coefficients \citep{Shen2012}.

\subsection{Visual Spectral Classification}
\label{sec:vis}

Alongside our automated broad-line detection and fitting approach, we have visually inspected each spectrum to identify AGN signatures and provide initial classifications. Sources displaying at least one broad emission line are classified as ``broad line'' objects. The remaining spectra, which lack clear AGN features, are either host-dominated or have low signal-to-noise ($S/N$). Given the varying quality of spectrophotometric calibrations, we do not attempt to identify AGN continuum features. 

In total, 174 sources exhibit AGN features, while 90 do not show obvious AGN-like characteristics, indicating either that they comprise host-dominated spectra or significant noise. These classifications are listed in Table~\ref{tab:mass}. However, the absence of detected AGN spectral features does not necessarily mean an AGN is not present. For instance, the source could be in a host-dominated phase due to variability, or the observed spectrum may not cover any broad emission lines at the given redshift.


\subsection{X-ray Properties}

\begin{figure}
\centering
\includegraphics[width=0.5\textwidth]{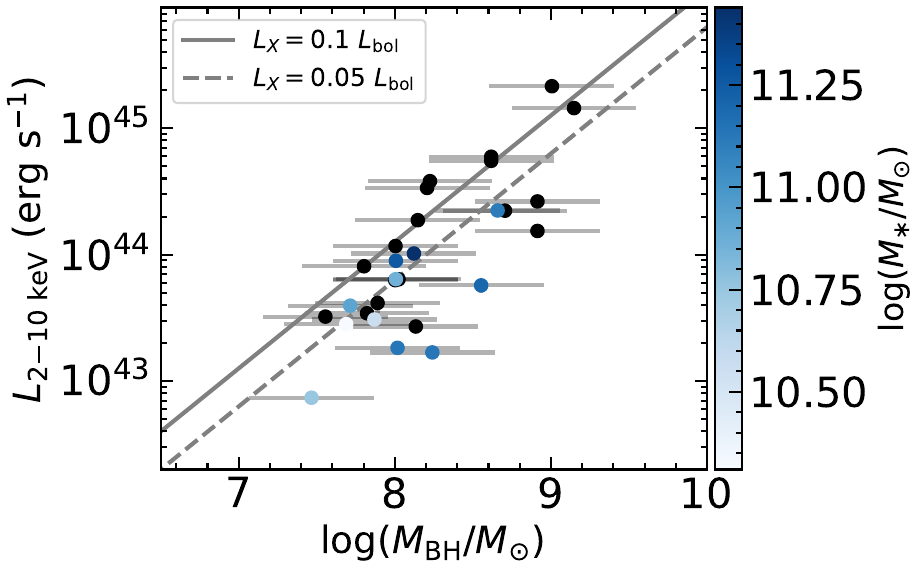}
\caption{The 2$-$10 keV X-ray luminosities plotted against our measured broad-line black hole masses. The points are color-coded by their stellar mass when it is considered reliable. An uncertainty of $0.4$ dex is assumed for the black hole masses. \label{fig:xray}}
\end{figure}

To estimate the rough accretion rates of our sources, we utilize matched X-ray data from the \emph{Chandra} Source Catalog and/or the XMM-Newton serendipitous survey catalog. The X-ray fluxes are converted to luminosities (uncorrected for absorption) using:  
\begin{equation}
L_{2-10\ {\rm keV}} = 4 \pi d^2\ (1+z)^{\Gamma-2} \frac{10^{2-\Gamma} - 2^{2-\Gamma}}{E_2^{2-\Gamma} - E_1^{2-\Gamma}} f_{E_1{-}E_2},
\end{equation}
where $f_{E_1{-}E_2}$ is the flux in keV given in the \emph{Chandra} Source Catalog or XMM-Newton serendipitous survey catalog between energies $E_1$ and $E_2$ in keV. We take $\Gamma=1.8$, which is typical for low-luminosity AGNs (e.g. \citealt{Ho2009}).

Figure~\ref{fig:xray} presents the X-ray luminosities plotted against our broad-line black hole masses. The X-ray luminosities, which are typically $\gtrsim 10^{43}$ erg s$^{-1}$, exceed the levels expected from X-ray binary populations \citep{Lehmer2019}. Applying a standard bolometric correction of $L_{\rm{bol}}/L_{2-10\ \rm{{keV}}} = 10$ \citep{Duras2020}, we find that the sources have a typical (median) Eddington ratio of  $\sim 0.05$, which is slightly lower than the median $\sim 0.1$ Eddington ratio reported by \citet{Suh2020}. Our estimated Eddington ratios and black hole masses align well with expectations, given the known correlation between these parameters and optical variability amplitude \citep[e.g.,][]{MacLeod2010}. This is consistent with the typical variability amplitude of DES-SN AGNs, which is around 0.1 magnitudes.

\section{$M_{\rm{BH}}-M_{\ast}$ relation} \label{sec:relation}

\begin{figure*}
\centering
\includegraphics[width=0.8\textwidth]{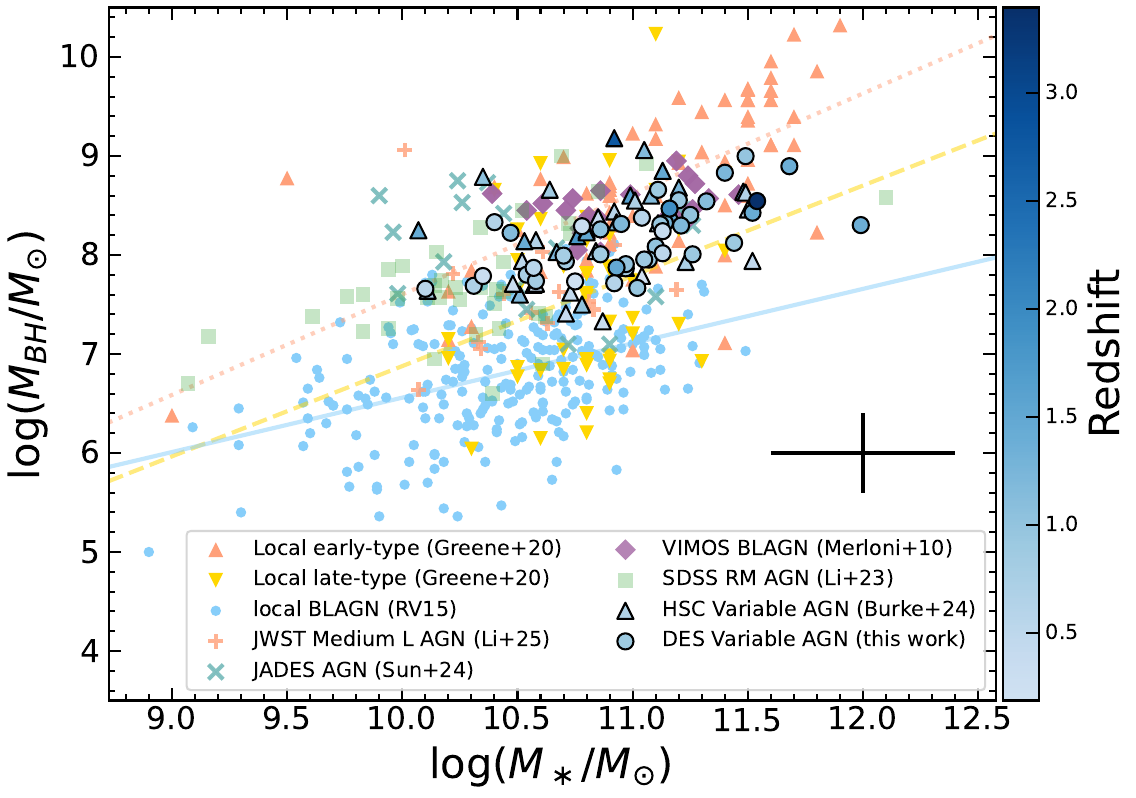}
\caption{$M_{\rm{BH}}-M_{\ast}$ relation for the DES-SN (circle symbols) and HSC-SSP (triangle symbols) AGNs with reliable BH mass estimates and reliable stellar masses from this work. Points are shaded by redshift. We assume uncertainties of $\sim 0.4$ dex on the BH masses. The typical uncertainties in our data are shown in the lower right corner. Data for AGN samples are overplotted with different symbols \citep{Li2025, Sun2024, Merloni2010, Burke2024hsc, Li2023}. The local $M_{\rm{BH}}-M_{\ast}$ relation for inactive elliptical galaxies is plotted as a red dotted line and for inactive spiral galaxies as a dashed yellow line \citep{Greene2020}. The local $M_{\rm{BH}}-M_{\ast}$ relation for AGNs is plotted as a solid blue line \citep{Reines2015}. 
\label{fig:relation}}
\end{figure*}

\begin{figure*}
\centering
\includegraphics[width=0.95\textwidth]{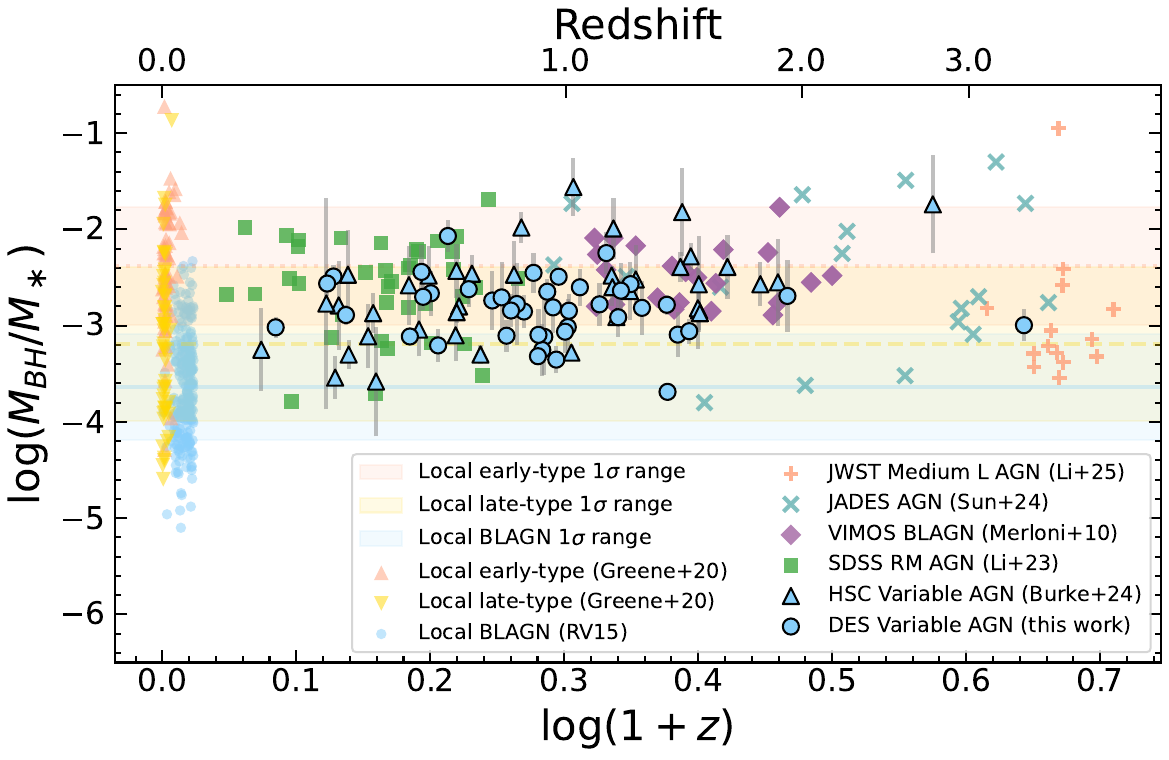}
\caption{Summary of $M_{\rm{BH}}-M_{\ast}$ versus redshift up to $z\sim4$. The DES-SN (circle symbols) and HSC-SSP (triangle symbols) AGNs with reliable BH mass estimates and reliable stellar masses are plotted in blue symbols with black borders. We include \citet{Li2025, Sun2024, Merloni2010, Burke2024hsc, Li2023} samples, along with local galaxies \citep{Greene2020} and local AGNs \citep{Reines2015} for comparison. The $1\sigma$ range of the local early-type galaxies, late-type galaxies, and AGNs are plotted as red, yellow, and blue spans with dotted, dashed, and solid lines, respectively. Our AGN samples have a larger redshift span, making it easier to compare with different samples.
\label{fig:mratioredshift}}
\end{figure*}

\begin{figure*}
\centering
\includegraphics[width=0.95\textwidth]{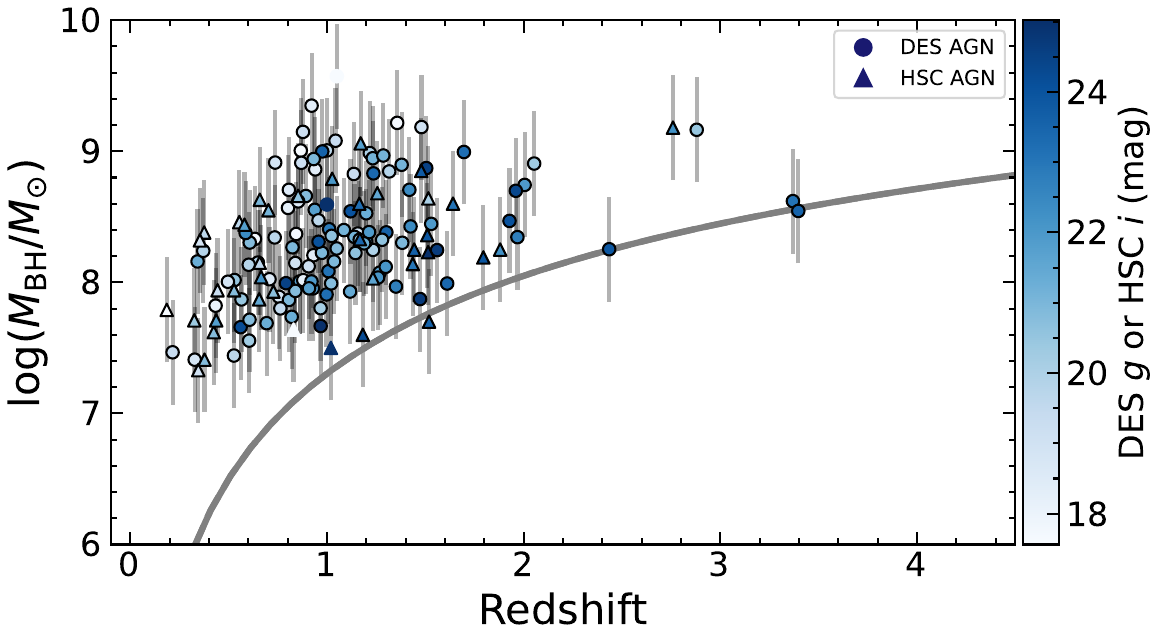}
\caption{Black hole mass as a function of redshift for DES-SN (circle symbols) and HSC-SSP (triangle symbols) variable AGNs with reliable broad-line BH mass estimates derived from broad-band SED fitting using \textsc{cigale}. Each AGN is color-shaded according to its DES $g$-band or HSC $i$-band apparent magnitude. The gray curve represents the theoretically predicted detection limit, which extends deeper than the BH mass detection limit. \label{fig:bhmassredshift}}
\end{figure*}

Figure~\ref{fig:relation} presents the $M_{\rm{BH}}-M_{\ast}$ relation for our sample of variability-selected AGNs with reliable stellar masses derived from \citet{Hartley2022} photometry and virial black hole masses measured from our spectroscopic database. For AGNs in the combined DES-SN and HSC-SSP sample with reliable black hole and stellar masses at $z\sim0.1-3.4$ (median redshift $\sim 0.8$), the relation aligns more closely with that of local inactive elliptical galaxies \citep{Greene2020} rather than local ($z<0.055$) AGNs \citep{Reines2015}. However, at lower redshifts, variability-selected AGNs appear to trend toward the local AGN relation. This trend, previously observed in HSC-SSP AGNs, is now confirmed with a larger sample size \citep{Burke2024hsc}. 

There is a notable absence of sources with very massive black holes ($M_{\rm{BH}} \gtrsim 10^9 M_\odot$) in our sample, compared to the local inactive elliptical galaxy population. We attribute this to the difficulty in obtaining reliable stellar masses for highly luminous AGNs with high AGN fractions, where host galaxy light is overwhelmed by the AGN emission.

At first glance, this result may suggest an evolution in the $M_{\rm{BH}}-M_{\ast}$ relation for AGNs with redshift. Such an evolution could be linked to changes in host star formation rates or black hole accretion activity, as both peak at $z \sim 2$ \citep{Zhuang2023nat}. However, \citet{Bongiorno2012} analyzed AGN activity and host star formation rates in an X-ray and optically selected AGN sample up to $z \sim 3$ in the COSMOS field, finding no strong evidence for a direct connection between AGN activity and host star formation. \citet{Hickox2014} suggested that the weak correlation between AGN properties and host SFR or stellar mass could be explained by the significantly shorter $\sim$ Myr scale timescales of AGN activity compared to the $\sim 100$ Myr timescales of star formation in galaxies. Additionally, weak trends seen in other studies \citep[e.g.,][]{Lutz2010, Bonfield2011} may be attributed to the large scatter in stellar mass and SFR estimates.

\section{Discussion} \label{sec:discussion}

\begin{figure*}
\centering
\includegraphics[width=0.99\textwidth]{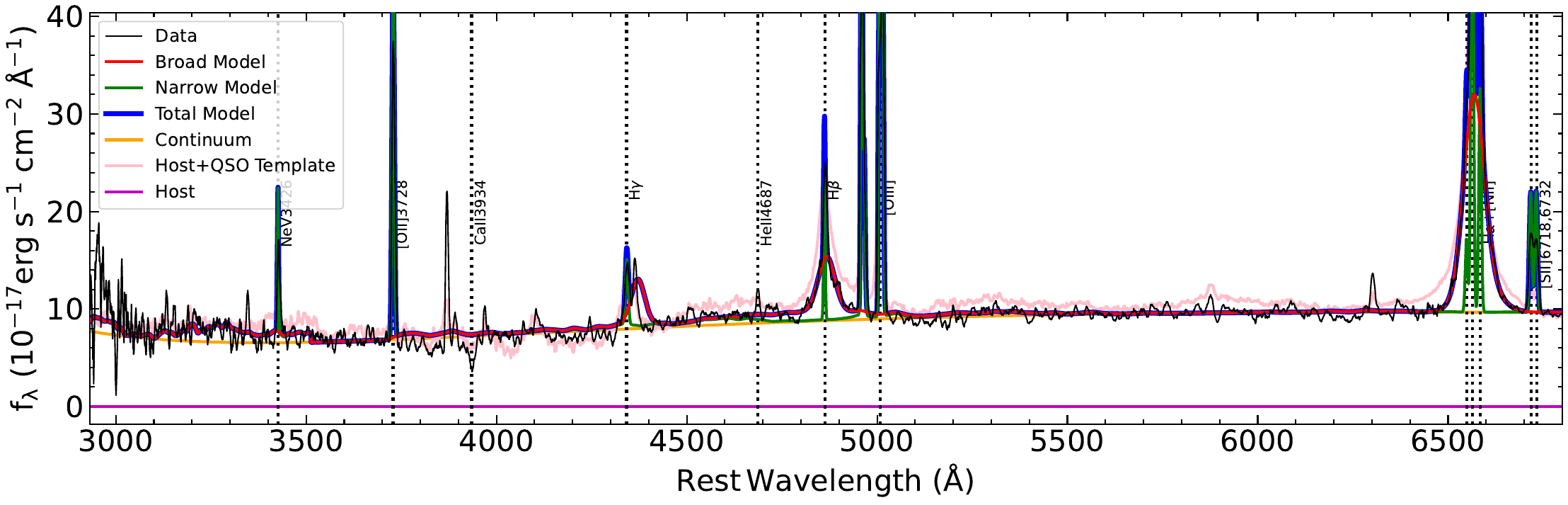}
\includegraphics[width=0.99\textwidth]{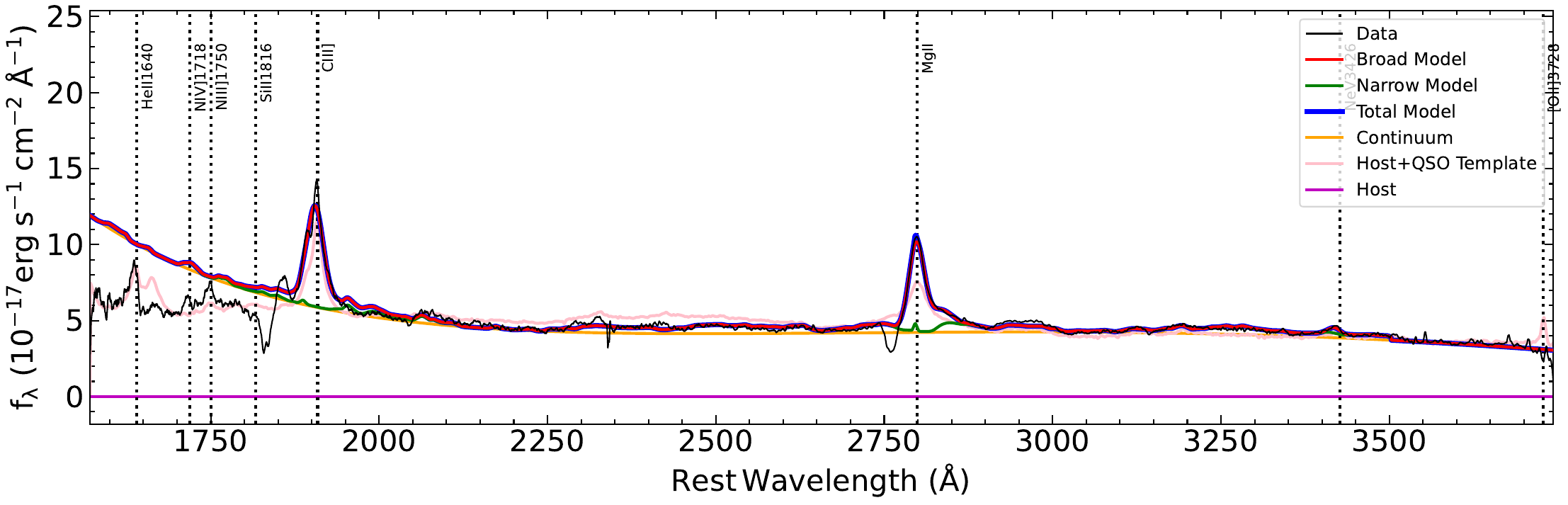}
\caption{Spectral fitting results for \citet{Burke2022des} ID 697 (upper panel) and ID 399 (bottom panel). The source in the upper panel is the lowest reliable stellar mass in our spectroscopic sample of $M_{\ast} \sim 10^{10} M_{\odot}$. The source in the lower panel is an example of a brighter AGN (showing blueshifted absorption features) with an unreliable stellar mass. The spectral data (black) is plotted with the model fitting components: polynomial+power-law continuum (orange), total line model flux (blue), broad line model flux (red), narrow line model flux (green) fitted host+QSO template (pink) and fitted host template (magenta). In both cases, the PCA-derived host fluxes are small. The resulting broad-line black hole mass for the source in the lower panel is $M_{\rm{BH}} \sim 10^{8.3} M_{\odot}$ at $z=1.383$. The black hole mass for the source in the upper panel is $M_{\rm{BH}} \sim 10^{7.5} M_{\odot}$ at $z=0.215$. \label{fig:spec}}
\end{figure*}

\subsection{Impact of Selection Biases}

Measurements of the $M_{\rm{BH}}-M_{\ast}$ relation are significantly affected by selection biases at both low and high redshifts \citep{Lauer2007,Shen2010,Shankar2019}, as discussed in \citet{Burke2024hsc}. Our spectroscopic sample includes spectra from various spectroscopic programs, each with different targeting criteria and spectral sensitivities. Additionally, we cannot reliably estimate stellar masses for optically luminous AGNs, making it challenging to quantify the impact of selection biases on our results.  

Selection biases have been shown to introduce a false evolution in the $M_{\rm{BH}}-M_{\ast}$ relation \citep{Lauer2007,Shen2010}. Since our sample is selected based on AGN activity, it is inherently biased toward more luminous AGNs and larger black hole masses at higher redshifts. However, previous studies have found a similar $M_{\rm{BH}}-M_{\ast}$ relation for X-ray and spectroscopically selected AGNs at comparable redshifts and higher AGN luminosities \citep{Merloni2010,Zhuang2023}. This suggests that the variable AGN population is unlikely to be probing a substantially different parameter space from the broader Type 1 AGN population.  

Quantifying the effects of selection biases on our sample would require modeling the detectable AGN population at these redshifts (e.g., \citealt{Pacucci2023,Li+2024}), but such an analysis depends on the true intrinsic scatter of the $M_{\rm{BH}}-M_{\ast}$ relation, which remains uncertain. Possible origins of the scatter in the $M_{\rm{BH}}-M_{\ast}$ relation include AGN variability, galaxy mergers, or SED fitting uncertainties. The findings in this paper, we acknowledge, should be considered within the context of these limitations. Detailed modeling of selection biases is beyond the scope of this work.

\subsection{Evolution in the $M_{\rm{BH}}-M_{\ast}$ relation?}

As noted by \citet{Burke2024hsc}, the recently proposed model of \citet{Pacucci2024} predicts that black holes at $z \sim 0.5 - 3$ should be $3 - 10$ times more massive than expected from the local AGN relation, which is broadly consistent with our findings. For example, the earlier models of black hole-galaxy coevolution, such as those by \citet{Wyithe2003} and \citet{Caplar2018}, also predict a redshift evolution of the $M_{\rm{BH}}-M_{\ast}$ relation that is attributed to self-regulated feedback which limits black hole growth before quenching occurs at a critical $M_{\rm{BH}}/M_{\ast}$ threshold. Additionally, \citet{Dattathri2024} proposes that the evolution of the mean relation is driven by the ratio of the black hole accretion rate to the star formation rate, $\dot{M}_{\rm{acc}}/{\dot{M}_\ast}$. 

If our results are not significantly affected by selection effects, they indicate a possible redshift evolution in the $M_{\rm{BH}}-M_{\ast}$ relation when compared to local AGNs \citep{Reines2015}. However, our results are generally consistent with no evolution beyond $z\sim.05$ (Figure~\ref{fig:mratioredshift}). However, the lack of any observed evolution beyond the local sample of \citet{Reines2015} --despite the significant redshift evolution in the cosmic star formation rate density-- seems to contradict these models. Perhaps differences in the sample selection of the \citet{Reines2015} makes these AGNs special in some way. Furthermore, it is possible that systematic errors in the black hole or stellar masses or model photometry are partly responsible for these discrepancies.



\section{Conclusions}  \label{sec:conclusion}

Using the sample of variable AGNs selected from the DES deep fields \citep{Burke2022des}, we obtained improved multiwavelength photometry from DES+VIDEO \citep{Hartley2022}, GALEX, WISE, and X-ray and used an improved SED fitting approach to estimate their stellar masses. We determined the reliability of the stellar mass estimates using the stellar emission strength at 1.2 $\mu$m, where the AGN emission is expected to be at a minimum \citep{Merloni2010,Burke2024hsc}. After constructing a database of publicly available spectra from the literature, we measured their virial black hole masses from the detected broad emission lines when $S/N$ permitted. We place our variable AGNs on the $M_{\rm{BH}}-M_{\ast}$ relation at $z\sim0.1-3.4$ (median redshift of $\sim 0.8$). The elevated $M_{\rm{BH}}/M_{\ast}$ ratios observed in variability-selected AGNs support scenarios where these SMBHs experienced early rapid growth compared to the \citet{Reines2015} local AGNs. These results, combined with the high occupation fraction in local dwarf galaxies \citep{Burke2025bhof}, are possibly consistent with a mixture of heavy seed formation channels.

This is a marked improvement over previous work \citep{Burke2022des}, because we have obtained homogenized broad-line BH mass estimates and improved the stellar mass estimates in this work. This allowed us to robustly place the sources on the $M_{\rm{BH}}-M_{\ast}$ relation. Compared to previous work in the COSMOS field \citep{Burke2024hsc}, this study fills the gap at $ z \lesssim 1.5$, forming a sample of variability-selected AGNs with well-sampled SEDs and BH masses at $ 0.05 \lesssim z \lesssim 3$. Our results concur with previous findings using other AGN samples of more massive black holes at a given stellar mass than the local AGN relation would suggest. These confirm our previous conclusion that AGNs selected from optical variability are not vastly different from samples of AGNs selected from broad lines at similar redshifts at fixed luminosity.

Using these results to assess the feasibility for such studies with capabilities of the LSST Rubin Observatory, we demonstrate that black holes with $M_{\rm{BH}} \sim 10^8 M_{\odot}$ are detectable out to at least $z\sim4$ in $M_{*} \sim 10^{11} M_{\odot}$ host galaxies using optical variability. Future work combining Rubin-selected AGNs and \textit{Euclid} or \textit{Roman Space Telescope} host galaxy imaging will increase the sample size of sources with reliable estimates of stellar masses from AGN$+$host decomposition, as joint analysis of of high-resolution and time-resolved imaging can yield improved SED extraction of AGN host galaxies \citep{Melchior2018,Ward2024}.


We have not investigated narrow emission line ratio diagnostics in this paper. Given the varying wavelength coverage and redshifts of the spectra, the majority of the spectra do not cover both the H$\beta$ and H$\alpha$ spectral complexes. The majority of the sources have AGN features in their spectra (either a broad line detection or Ne V emission line), but a significant fraction are either too noisy or host-dominated to detect any obvious AGN features. Some of the latter could be false positives (i.e., non-AGN galaxy or transient interlopers). This work highlights the challenges of obtaining sufficient spectroscopy to investigate low luminosity AGNs at the redshifts that LSST Rubin will unveil. 

The large scatter in the $M_{\rm{BH}}-M_{\ast}$ relation likely reflects the diversity of growth channels of AGN host galaxies and/or the mismatch between growth timescales of the black hole and star formation (e.g., \citealt{Hickox2014}). This and relatively low redshift of the sample, significantly later than the principal seeding epoch, expected to be at $z>10$ makes it difficult to place direct meaningful constraints on SMBH seeding pathways. However, estimates of the occupation fraction in low mass galaxies have been shown to be a reliable proxy of high redshift seeding from previous modeling work \cite{Volonteri2009,Ricarte2018}. Observational studies such as this one and future work spanning an even wider range of redshifts, host galaxy masses, and with diverse selection techniques are nevertheless important to anchor these models of SMBH seeding and feedback over cosmic time (for e.g. \citet{Volonteri2009,Ricarte2018}).

\section*{Acknowledgements}
This research has made use of the SIMBAD database \citep{Wenger2000}, operated at CDS, Strasbourg, France. This research made use of \textsc{astroquery} \citep{Ginsburg2019}, SIMBAD \citep{Wenger2000}, and Vizier \citep{vizier2000}. 

C.J.B. is supported by an NSF Astronomy and Astrophysics Postdoctoral Fellowship under award AST-2303803. This material is based upon work supported by the National Science Foundation under Award No. 2303803. This research award to NSF is partially funded by a generous gift of Charles Simonyi to the NSF Division of Astronomical Sciences. The award is made in recognition of significant contributions to Rubin Observatory’s Legacy Survey of Space and Time. 
Y.L. and X.L. acknowledge support from the University of Illinois Campus Research Board Award RB22065 and NSF grant AST-2308077. P.N. acknowledges support from the Gordon and Betty Moore Foundation and the John Templeton Foundation that fund the Black Hole Initiative (BHI) at Harvard University where she serves as one of the PIs.



\appendix

\section{DATA AVAILABILITY} \label{sec:availibility}


\noindent The publicly-available spectra were collected from the following web pages:
\vspace{-2mm}
\begin{enumerate}
    \item CDF-S faint X-ray: \url{https://member.ipmu.jp/john.silverman/CDFS_data.html} \& \\ \url{https://www.eso.org/~vmainier/cdfs_pub/} 
    \item Spitzer/IRS ATLAS: \url{https://www.denebola.org/atlas/}
    \item VVDS: \url{http://cesam.lam.fr/vvds}
    \item VIPERS: \url{http://vipers.inaf.it/ }
    \item GAMA: \url{https://www.gama-survey.org}
    \item OzDES: \url{https://docs.datacentral.org.au/ozdes/overview/ozdes-data-release/}
    \item SDSS: \url{https://dr18.sdss.org/optical/plate/search}
    \item DESI: \url{https://datalab.noirlab.edu/desi/access.php}
\end{enumerate}

\section{RELIABILITY OF SPECTRAL CONTINUUM MEASUREMENTS} \label{sec:conti}

\begin{figure}
\centering
\includegraphics[width=0.5\textwidth]{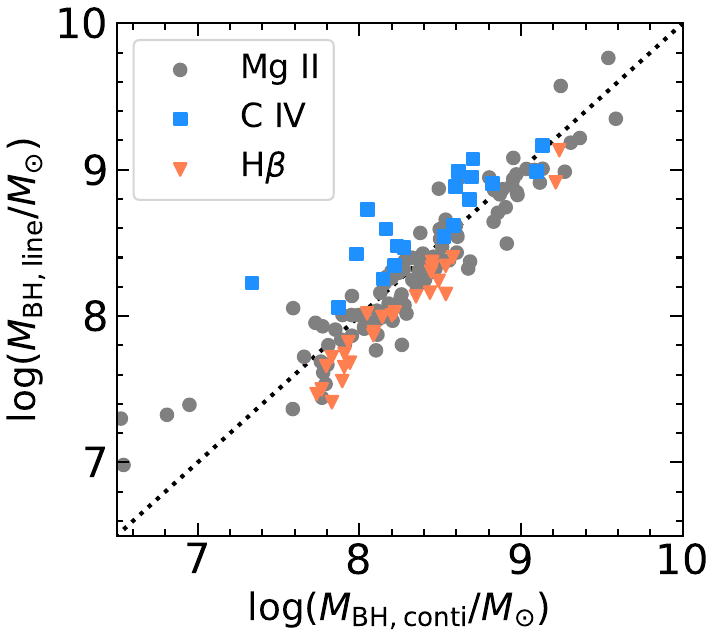}
\caption{Black hole masses estimated from the broad-line luminosity versus ($y$ axis) AGN continuum luminosity ($x$ axis). The black hole masses from the AGN continuum luminosity are susceptible to a few catastrophically underestimated masses when the AGN continuum is not well constrained. \label{fig:conti}}
\end{figure}

Spectra of AGNs with low bolometric luminosities of $L_{\rm{bol}} \lesssim 10^{45}$ erg s$^{-1}$ have a significant contribution from the underlying host galaxy \citep{Shen2011,Kimura2020}. Reliably constraining the continuum luminosity from the AGN is an essential step for obtaining a virial black hole mass estimate. This is because the virial prescriptions use the AGN continuum luminosity as a proxy for the BLR luminosity \citep{Shen2011}. PCA decomposition can constrain the quasar continuum for sources below $L_{\rm{bol}} \sim 10^{45}$ erg s$^{-1}$ if the spectra are of sufficient quality \citep{Ren2024}. However, the quality of the spectra in our sample (calibration and $S/N$) varies considerably depending on the instrument and the spectral reduction/calibration. Following our previous work \citep{Burke2024hsc}, we test whether the quasar continuum luminosities are well-constrained by computed virial black hole masses using both the broad-line luminosity (Equation~\ref{eq:BHmass}) or continuum luminosity approaches using the equation:
\begin{equation}
\begin{split}
    \log{\left(\frac{M_{\rm{BH}} }{M_{\odot}} \right)} = a + b \log{ \left( \frac{\lambda L_\lambda}{10^{44}\ \rm{ erg\ s}^{-1}} \right) } + 2 \log{ \left( \frac{\rm{FWHM}_{\rm{br}}}{\rm{ km\ s}^{-1}} \right) }
\end{split}
\end{equation}
where $\lambda L_\lambda$ and FWHM$_{\rm{br}}$ are the continuum luminosity and broad-line full-width-at-half-maximum (FWHM). We adopt the calibrations \citep{Vestergaard2006} used in \citet{Shen2011}:
\begin{equation}
    (a, b) = (0.910, 0.50), \quad \rm H\beta
\end{equation}
\vspace{-6mm}
\begin{equation}
    (a, b) = (0.740, 0.62), \quad \rm Mg\ II
\end{equation}
\vspace{-6mm}
\begin{equation}
    (a, b) = (0.660, 0.53), \quad \rm C\ IV.
\end{equation}

These single-epoch relations have an intrinsic scatter of $\sim0.4$ dex in black hole mass. If the AGN continuum luminosities are well-constrained, we expect the two approaches to yield consistent results within systematic uncertainties between the two prescriptions. For our sample, we found that two prescriptions typically yield similar results. However, the continuum-based black hole masses are often underestimated at the low-mass end when the AGN continuum is not well constrained, as shown in Figure~\ref{fig:conti}. Therefore, as in \citet{Burke2024hsc}, we adopt black hole masses estimated from the broad line luminosity throughout this work from Equation~\ref{eq:BHmass}.

\section{IMPROVEMENT OF STELLAR MASS ESTIMATES} \label{sec:dchi2}

\begin{figure}
\centering
\includegraphics[width=0.6\textwidth]{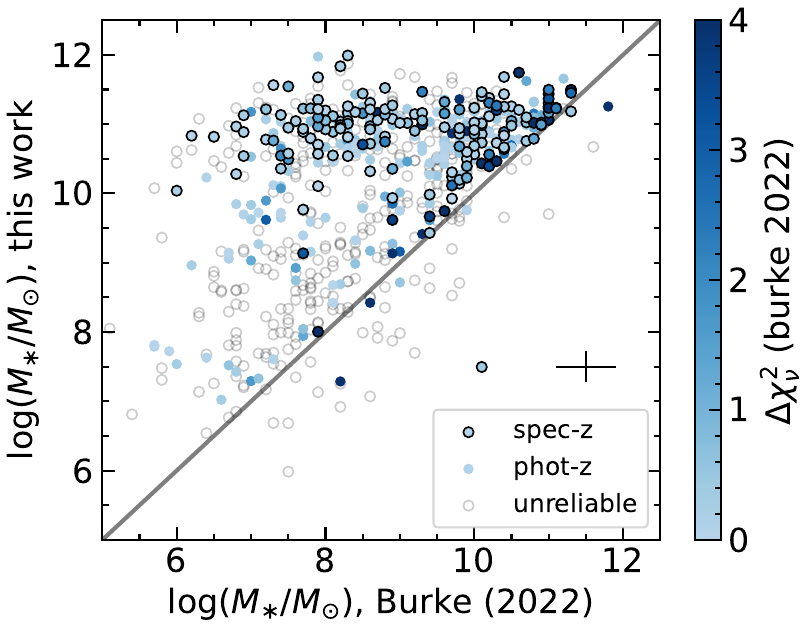}
\caption{Stellar masses estimated using our improved approach (this work) vs. the stellar masses calculated by \citet{Burke2022des}. The gray circles are unreliable stellar masses ($\chi^2 > 5$ or SF excess $> 1.2$ (same as Figure \ref{fig:bhmassredshift}). The circles with solid borders are sources with reliable stellar masses with spectroscopic redshifts. The circles without borders are reliable stellar masses with photometric redshifts. The circles are colored by the improvement in fit quality measure $\Delta \chi_\nu^2$ as calculated by \citet{Burke2022des}. The bluer colors are more reliable stellar masses, as determined by \citet{Burke2022des}. A typical error bar is shown on the right. More of our stellar masses are reliable using the improved approach. A $y=x$ line is plotted in gray. }\label{fig:dchi2}
\end{figure}

We deem our stellar masses reliable if the stellar emission dominates over the AGN emission at 1.2 $\mu$m, using the criteria $SF_\text{ex} > 1.2$ \citep{Burke2024hsc}. In cases where the AGN emission dominates, the SED-fitting-derived stellar masses are likely to have a large variance. To investigate this improvement, we plot the stellar mass calculated from this work and the stellar masses from \citet{Burke2022des} in Figure \ref{fig:dchi2}, where we use the $SF_\text{ex} > 1.2$ criteria to ensure that our stellar masses are reliable. The figure indicates that \citet{Burke2022des} generally underestimated the stellar masses, and sometimes by up to 4 orders of magnitude. However, most of the blue points (stellar masses considered reliable by \citet{Burke2022des}) lie on the $y=x$ line. It is reassuring that the \citet{Burke2022des} reliable stellar masses are generally consistent with ours. However, we conclude that more of our stellar masses are reliable. We attribute this primarily to the use of more photometry (including Chandra and XMM X-ray, Galex ultraviolet, and Spitzer infrared data) in our work.

\bibliography{example}{}
\bibliographystyle{aasjournal}



\end{document}